\documentclass[11pt,a4paper]{article}
\usepackage[margin=1in]{geometry}
\usepackage[english]{babel}
\usepackage{amsmath,amsthm}
\usepackage{amsfonts}
\usepackage{graphicx}
\usepackage{subfig}
\usepackage{float}
\usepackage{authblk}
\newcommand{\te}[1]{\mbox{\scriptsize $#1$}}

\begin{document}
\title{Test of the no-signaling principle in the Hensen\\loophole-free CHSH experiment}%
\author[1]{Guillaume ADENIER\thanks{guillaume.adenier@gmail.com}}
\author[2]{Andrei Yu. KHRENNIKOV\thanks{andrei.khrennikov@lnu.se}}
\affil[1]{INRiM, Strada delle Cacce 91, Torino 10135, Italy}
\affil[2]{International Center for Mathematical Modeling\\in Physics, Engineering, Economics, and Cognitive Science\\Linnaeus University, V\"{a}xj\"{o}, Sweden}
\maketitle
\begin{abstract}
We analyze the data from the loophole-free CHSH experiment performed by Hensen \emph{et al}, and show that it is actually not exempt of an important loophole. By increasing the size of the sample of event-ready detections, one can exhibit in the experimental data a violation of the no-signaling principle with a statistical significance at least similar to that of the reported violation of the CHSH inequality, if not stronger.
\end{abstract}

\section{Introduction}
In special relativity, two events taking place at distant locations are space-like separated when there is not enough time for any causal influence to propagate (at the speed of light) between the two events.

Consider Alice and Bob, performing measurements at distant laboratories A and B in order to implement a Bell-CHSH test of local realism \cite{Bell,CHSH,Aspect,Weihs,Genovese}. We label their measurement settings as $a$ and $b$ respectively, and their measurement results as $x$ and $y$, with $x,y\in{\{\pm1\}}$. If the two events $(a,x)$ and $(b,y)$ are space-like separated, then Bob's outcome $y$ cannot reveal any information about Alice's input $a$, and the other way around. The no-signaling principle is just that: local measurements performed at one station cannot be used to send a signal to the other station, when the stations are space-like separated.

In spite of being nonlocal, quantum theory is not at variance with this principle \cite{GRW}. The no-signaling principle is in fact recognized as an essential characteristic of quantum theory \cite{Brunner}. The predictions of quantum theory being intrinsically random, the nonlocal behavior is fundamentally masked by randomness at the local level. It cannot be used to transmit a signal at a distance.

Since the existence of a signalling channel between Alice and Bob could be exploited to mimic nonlocal correlations with a local model, testing the no-signaling principle is especially important in the context of a claim of an experimental violation of local realism. While it used to remain largely untested in the literature, with the exception perhaps of our analysis \cite{Adenier} of the 1997-1998 Innsbruck experiment \cite{Weihs}, and that of De Raedt \emph{et al} \cite{DeRaedt}, it is now becoming standard practice to test the no-signalling principle jointly with tests of local realism \cite{GiustinaSup,ShalmSup,Rosenfeld}.

In this line of work, we propose to perform an alternative analysis of the Hensen \emph{et al} experiment \cite{Hensen}, in which the violation of local realism is investigated jointly with testing for possible violation of the no-signaling principle.\footnote{This question was also raised by Bednorz \cite{Bednorz}, but with a rather too succinct analysis to be able to conclude unequivocally.}

\section{Theory}

\subsection{Testing local realism}

We briefly recall the mathematical tools and notations used to perform a CHSH test of local realism \cite{Bell,CHSH,Aspect,Weihs,Genovese}.

\subsubsection{Bell-CHSH inequality}

The joint probabilities to observe events $x$ and $y$ given the settings $a$ and $b$ are written $p(xy|ab)$.

With measurement settings $a$ and $b$, the correlation function is written:
\begin{equation}\label{correlation}
  E_{ab}=p(\te{++}|ab)-p(\te{+-}|ab)-p(\te{-+}|ab)+p(\te{--}|ab)
\end{equation}

In order to put local realism \cite{Bell} to an experimental test, the CHSH inequality \cite{CHSH} has long been the favorite weapon of choice \cite{Aspect,Weihs}:
\begin{equation}\label{CHSH}
  S_\textrm{CHSH}=|E_{00}+E_{01}+E_{10}-E_{11}|\leq 2
\end{equation}

\subsubsection{Statistics with finite sample}

An experimental implementation of a Bell-CHSH test provides a set of number of counts $N^{xy}_{ab}$: the number of times that events $x$ and $y$ are jointly registered when the settings are $a$ and $b$. The joint probabilities $p(xy|ab)$ are not directly accessible  to the experimenter and must be estimated as the ratio of the corresponding number of counts $N^{xy}_{ab}$ over the total number of counts:
\begin{equation}\label{freq}
  p(xy|ab)=\frac{N^{xy}_{ab}}{N^{++}_{ab}+N^{+-}_{ab}+N^{-+}_{ab}+N^{--}_{ab}}
\end{equation}

The random character of the counts and their finite size mean that an error in the estimation of the joint probabilities has to be taken into account, as it could very well propagate to the inequality being tested, possibly leading to an incorrect interpretation of the result.

\subsubsection{Estimation of the statistical error}

In the experiment that most particularly interests us, the  Hensen \emph{et al} \cite{Hensen} experiment, only a single measured count of each type $N^{xy}_{ab}$ is available. This is so for practical reasons due to the extremely low rate of pair productions: a valid trial event is registered about every hour or so, so that it takes days and weeks to record a sufficient number of counts.  Because of this limitation, the statistical error being made cannot be measured by calculating the variance of a series of counts. Instead, it has to be estimated with additional hypotheses. This is quite standard for this type of experiment, and it is arguably a trivial statistical tool that need not be recalled. Nevertheless, the low number of counts obtained in the experiment renders this problem much more crucial than it usually is, so that we prefer to write it down explicitly.

We first make the assumption that the number of counts $N^{xy}_{ab}$ are independent random variables. Let
$$N_1,N_2,...,N_i,...,N_m$$
be $m$ independent random variables with their corresponding standard deviations
$$\sigma_1,\sigma_2,...,\sigma_i,...,\sigma_m$$

Let
$$f(N_1,N_2,...,N_i,...,N_m)$$
be a function of these random variables. The propagation of standard deviations to this function of random variables is calculated as \cite{Taylor}:

\begin{equation}\label{propag}
  \sigma_f^2=\sum_{j=1}^m \big(\frac{\partial f}{\partial N_j}\big)^2\sigma_j^2
\end{equation}

We now assume that the statistics governing each number of counts is Gaussian, which means that the associated standard deviation $\sigma_i$ is the square root of each number $N_i$:
\begin{equation}\label{stddev}
  \sigma_i=\sqrt{N_i}
\end{equation}

Under theses assumptions, one can calculate the standard deviations associated with any measured frequencies of the type Eq.~(\ref{freq}), which in turn can be used to calculate the standard deviation associated with any sum and differences of these frequencies, as their errors should simply be added in quadrature \cite{Taylor}.

For instance, using Eqs.~(\ref{correlation}), (\ref{freq}), (\ref{propag}) and (\ref{stddev}), it is straightforward to show that the standard deviation associated with the measured correlation function $E_{ab}$ is:
\begin{equation}\label{sigmacorr}
  \sigma_{E_{ab}}^2=\frac{4(N_{ab}^{++}+N_{ab}^{--})(N_{ab}^{+-}+N_{ab}^{-+})}{(N_{ab}^{++}+N_{ab}^{+-}+N_{ab}^{-+}+N_{ab}^{--})^3}
\end{equation}

Armed with Eqs.~(\ref{correlation}), (\ref{CHSH}), and (\ref{freq}), one can put local realism to an experimental test, and estimate the magnitude of the error with Eq.~(\ref{sigmacorr}), in what can be coined the \emph{conventional} analysis \cite{Hensen}. Note that Hensen \emph{et al} used as well a more complicated analysis to take into account possible memory effects and predictability of measurement settings \cite{HensenSup}, but we prefer to stick with the simpler conventional analysis as it does not fundamentally change the results at hand (see Section \ref{relevance}).

\subsection{Testing the No-Signalling Principle}

The no-signaling principle that we want to test here can be expressed mathematically in terms of probabilities. One can write explicitly the no-signaling constraints to guarantee that Alice and Bob cannot use local measurements on quantum states to signal one another.

\subsubsection{Marginal probabilities and No-signaling equalities}

In a no-signalling experimental framework, Alice's marginal probabilities, which are the sum of joint probabilities over all possible results at Bob's laboratory, must be independent of the measurement performed by Bob. The no-signaling constraints \cite{Brunner} are therefore written for Alice as follow:
\begin{equation}\label{margA}
  p_\textrm{A}(x|a)\equiv p_\textrm{A}(x|ab)=\sum_{y\in\{\pm1\}} p(xy|ab)
\end{equation}
for all $x$, $a$, $b$. This expression conveys explicitly that Alice's marginal probabilities are independent on the measurement settings $b$ chosen by Bob, and that this must hold regardless of Alice's measurement settings $a$ and outcome $x$.

Similarly, for Bob we write:
\begin{equation}\label{margB}
  p_\textrm{B}(y|b)\equiv p_\textrm{B}(y|ab)=\sum_{x\in\{\pm1\}} p(xy|ab)
\end{equation}
for all $y$, $a$, $b$.

We are interested in testing the no-signaling principle for actual implementation of Bell-CHSH tests, in particular for the Hensen loophole-free Bell experiment \cite{Hensen,HensenSup}. In a CHSH experiment, it is sufficient to perform measurements with two measurements settings 0 and 1 at each laboratory, as written in Eq.~(\ref{correlation}). Since the data at hand deals only with two such measurement settings 0 and 1, we thus write explicitly Alice's no-signaling constraints for the +1 detections with these settings:
\begin{align}
    p_\textrm{A}(\te{+}|0)&= p_\textrm{A}(\te{+}|00)=p_\textrm{A}(\te{+}|01), \\
    p_\textrm{A}(\te{+}|1)&= p_\textrm{A}(\te{+}|10)=p_\textrm{A}(\te{+}|11).
\end{align}
These equations mean that Alice's marginal probabilities do not depend in any way in the measurement settings chosen by Bob (0 or 1).

We can rewrite them as no-signaling equalities in the direction from Bob to Alice:
\begin{align}\label{NoSignBtoA0}
  S_{\textnormal{B}\rightarrow\textnormal{A}_0}&=p_\textnormal{A}(\te{+}|00)-p_\textnormal{A}(\te{+}|01)=0,\\
  S_{\textnormal{B}\rightarrow\textnormal{A}_1}&=p_\textnormal{A}(\te{+}|10)-p_\textnormal{A}(\te{+}|11)=0.
  \label{NoSignBtoA1}
\end{align}

Similarly, in the direction from Alice to Bob, we write:
\begin{align}\label{NoSignAtoB0}
  S_{\textnormal{A}\rightarrow\textnormal{B}_0}&=p_\textnormal{B}(\te{+}|00)-p_\textnormal{B}(\te{+}|10)=0\\
  S_{\textnormal{A}\rightarrow\textnormal{B}_1}&=p_\textnormal{B}(\te{+}|01)-p_\textnormal{B}(\te{+}|11)=0
  \label{NoSignAtoB1}
\end{align}

We could write similar equation using the marginal probabilities for the -1 results, but using the fact that the marginal probabilities for +1 and -1 are adding up to unity at each location, we would end up with the exact same equations.

Admittedly, these no-signaling equalities have an air of triviality, since they are simple and always fulfilled by quantum theory as well as by any local realist theory \cite{Brunner}, which is probably why they were not tested until recently \cite{Adenier,DeRaedt,GiustinaSup,ShalmSup,Rosenfeld}. They are nevertheless essential in a test of local realism since their failure would invalidate the test altogether. In the case of ideal detectors, they can be expressed as function of the same joint probabilities $p(xy|ab)$ that are used to calculate the CHSH function, using the righthand side of Eqs.~(\ref{margA}) and (\ref{margB}), and can therefore be tested with the same set of experimental data required in a test of local realism.

\subsubsection{Estimation of the statistical error on the marginal probabilities}

For Alice, the standard deviation on her marginal probabilities is calculated using
using Eqs.~(\ref{freq}), (\ref{propag}), (\ref{stddev}) and (\ref{margA}) as:
\begin{equation}\label{sigmaA}
  \sigma^2_{A^+_{ab}}=\frac{
(N_{ab}^{++}+N_{ab}^{+-})(N_{ab}^{-+}+N_{ab}^{--})
}{(N_{ab}^{++}+N_{ab}^{+-}+N_{ab}^{-+}+N_{ab}^{--})^3}.
\end{equation}

Similarly, for Bob, using Eq.~(\ref{margB}), we get:
\begin{equation}\label{sigmaB}
  \sigma^2_{B^+_{ab}}=\frac{
(N_{ab}^{++}+N_{ab}^{-+})(N_{ab}^{+-}+N_{ab}^{--})
}{(N_{ab}^{++}+N_{ab}^{+-}+N_{ab}^{-+}+N_{ab}^{--})^3}.
\end{equation}

Armed with Eqs.~(\ref{freq}), (\ref{margA}), (\ref{margB}), (\ref{sigmaA})  and (\ref{sigmaB}), one can put the no-signaling principle to an experimental test and estimate the magnitude of the error with the same conventional analysis as in the CHSH case.

\section{Application to Hensen \emph{et al} experiment}

\subsection{The experiment in a nutshell}

\begin{figure}[htp]
    \centering
    \includegraphics[width=0.8\linewidth,height=0.8\textheight,keepaspectratio]{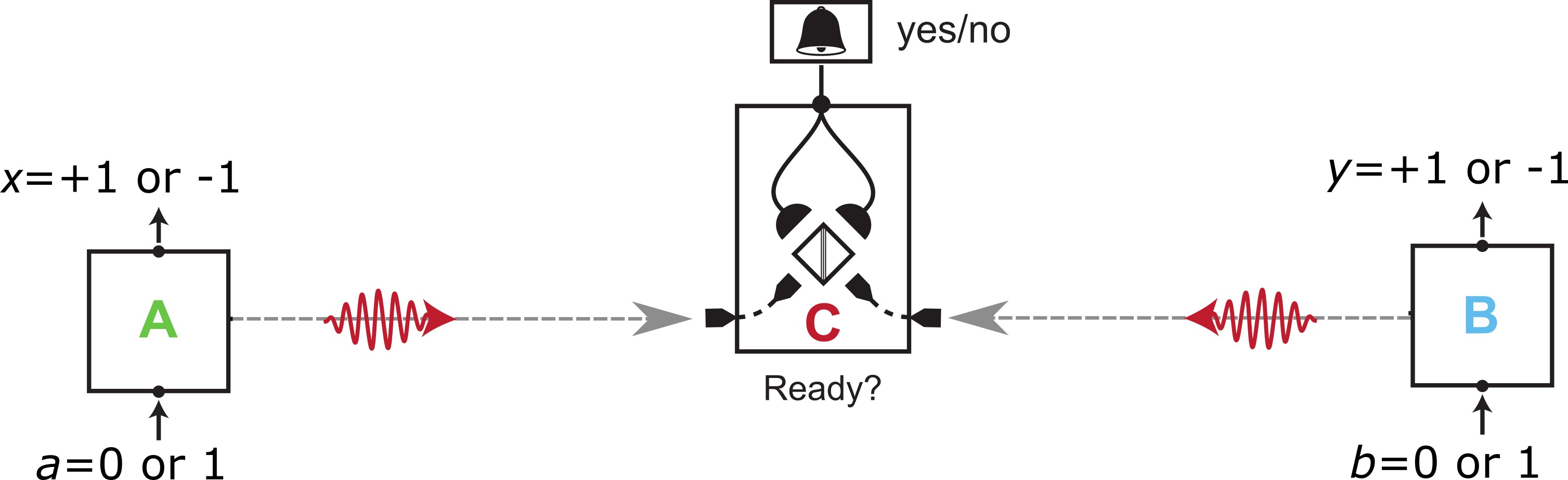}
    \caption{The experimental setup. Alice and Bob perform space-like separated measurements at stations A and B with settings $a$ and $b$ respectively, and results $x$ and $y$. A successful entanglement between A and B is marked by a coincidence detection at location C (event-ready protocol).}
    \label{fig:setup}
\end{figure}

Alice and Bob are located at distant laboratories. They each own a single nitrogen vacancy (NV) centre electron spin in diamond. They prepare these NV centres with laser excitation pulses, rotate those spins with microwaves (see Fig.\ref{fig:setup}).

Consequently to this preparation, photons emitted by the NV centres  are sent to overlap at a distant beam splitter located a C, roughly midway between Alice and Bob. A coincidence detection at the output ports of this beam splitter is used as an event-ready signal. It indicates successful photon entanglement, and therefore, by virtue of entanglement swapping, a successful entanglement between the NV centres \cite{Hensen,HensenSup}.

The measurements settings $a$ and $b$ are chosen randomly. The switching of the measurement bases and the actual measurements themselves are performed fast enough such that they are space-like separated and that no causal influence can travel between the two, which rules out the locality loophole in this experiment.

During the readout, the measurement always provides a result: either \emph{at least one photon is detected} (+1), which is likely to happen when the NV centre is in a bright spin state and therefore emit many photons, or \emph{no photon is detected} (-1), which corresponds the NV centre remaining dark. The measurement is therefore 100$\%$ efficient since these results are mutually exclusive. Errors are possible, for instance a dark state could be mistaken for a bright state because of a dark count occurring in the photon detector, but it would only affect the fidelity of the measurement, and not its efficiency. The detection loophole is thus ruled out as well in this experiment.

Another advantage is that the perfect detection efficiency makes it possible to test the no-signalling principle directly. With low efficiency detectors, an apparent violation of the no-signaling principle can indeed easily appear when the efficiencies at one measuring station (Alice or Bob) are unbalanced for the two possible results $\pm1$, so that additional hypotheses are required to test this principle \cite{Adenier}. Here, the perfect detection efficiency circumvents this caveat entirely.

\subsection{Event-ready sampling}

Although the detection loophole is ruled out by the design of the experiment, there is nevertheless some form of sampling involved in the analysis. It occurs at the event-ready level, when one has to chose the parameters that determine whether the NV centres at the measuring stations A and B are successfully entangled and therefore ready for a measurement.

It should be noted that this type of event-ready sampling is in principle perfectly fine in the context of a test of local realism. One should be free to decide beforehand what is an event-ready detection and what is not. It is indeed in principle very different than the sampling occurring due to low detection efficiency in photon twin-beam experiments \cite{Aspect,Weihs}. The point is that this sampling happens here \emph{before} the measurement settings are chosen, so that it is by design a fair sampling with respect to the settings, since it cannot possibly depend on undetermined measurement settings. By contrast, in an experiment with low detection efficiency the possibility that the sampling happens after the measurement settings are chosen cannot be ruled out. This is especially true with avalanche photon detectors as they are necessarily located at the very end of the path followed by the photons, therefore after the measurement settings are chosen, and have typically low quantum efficiency.

In the case of Hensen \emph{et al} experiment, the necessity of this event-ready sampling comes from the fact that many event-ready detections are in fact due to unwanted reflections of the laser excitation pulses, instead of coming from photons emitted by the NV centres \cite{HensenSup}. While a coincidence from photons emitted by the NV centres transfers the entanglement to the NV centres themselves by entanglement swapping, this is of course not the case when the coincidence originates from unwanted reflections of the laser excitation pulses. These unwanted event-ready events are therefore bound to deteriorate the observed correlation, and must be filtered out as best as possible in order to be able to observe a violation of the CHSH inequality.

The event-ready signal is a coincidence detection occurring within a precisely defined window after the arrival time of a sync pulse. The unwanted event-ready detections due to reflections of the laser excitation pulses can be filtered by delaying the start time of the coincidence window, but since the NV centre emissions and the laser excitation reflections partially overlap in time, as can be seen from Figure S2 in the Supplementary Information to the \emph{Nature} article \cite{HensenSup}, one cannot filter out one without filtering the other as well, at least partially. The start time of the window  must therefore be chosen with care in order to observe a violation of the CHSH inequality. Too early, and the photons coming from the reflections of the laser excitation pulses become predominant. Too late, and not enough valid photons from the NV centres are picked up, so that noise becomes predominant \cite{HensenSup}.

\subsubsection{Varying the sample size of event-ready detections}

A very interesting feature of the data provided by Hensen \emph{et al} is that the parameters determining the event-ready window are not fixed. They can be changed \emph{a posteriori} during the analysis, since all photon detection times at location C are recorded in the data. This feature has already been used by Hensen \emph{et al} to investigate the effect of choosing different event-ready samples in post-processing, in particular in Fig.S3 in the Supplementary Information to the \emph{Nature} article \cite{HensenSup}. In this respect, we follow their footsteps.

We will show below that changing these parameters to an even larger range reveal some valuable information that would not have been available otherwise, in particular to test the no-signaling principle.

\begin{figure}[htp]
    \centering
    \includegraphics[width=0.8\linewidth,height=0.8\textheight,keepaspectratio]{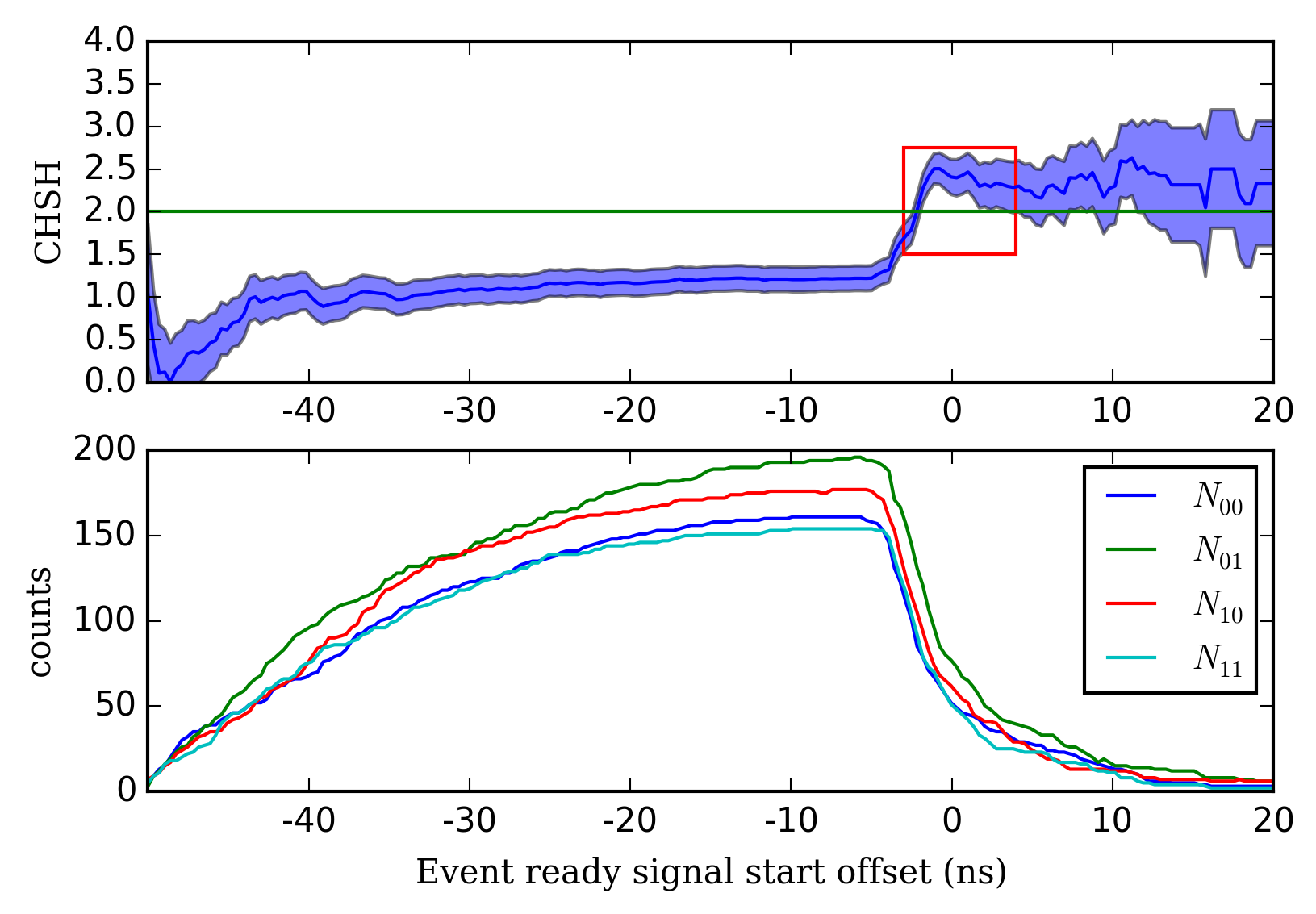}
    \caption{CHSH parameter $S$ and number of trials as a function of the window start for the event-ready sampling. The time-offset origin at 0 corresponds to the value chosen by Hensen \emph{et al}. The confidence region shown is one standard deviation, calculated according to the conventional analysis. The red rectangle in the CHSH plot highlights the range of the plot in Fig.S3 in Ref.\cite{HensenSup}.}
    \label{fig:CHSH}
\end{figure}

As hinted above, the choice of the coincidence-window setting determining the event-ready sample is not conservative for the CHSH parameter: if one chooses different values than the one used by Hensen \emph{et al} \cite{HensenSup}, then the size of the sample of trials changes quickly, as well as the value of the CHSH parameter $S$. This is illustrated on Fig.~\ref{fig:CHSH}, where the starting time of the even-ready window with respect to the synchronisation signal is varied over a large range, that is, larger than in Fig.S3 in the Supplementary Information to the \emph{Nature} article \cite{HensenSup}. We chose a range of possible offsets from -50 ns to 20 ns, instead of from -3 ns to 4 ns in \cite{HensenSup}. The reason for choosing such a large range will be apparent as we proceed with testing the no-signaling principle.

All other parameters are kept as set by Hensen \emph{et al}. The offset 0 corresponds to the value chosen beforehand by the experimenters, which yields the violation of $S=2.422 \pm 0.204$
reported in \cite{Hensen,HensenSup}. Negative offset means that more unwanted events coming from reflections of the laser excitation pulses are included in the event-ready sample. Positive offset corresponds to moving further away the window from those unwanted event-ready events, which also means that less and less valid photons from the NV centres are picked up, so that noise becomes predominant.

\subsection{Testing the no-signaling principle}

In order to test the no-signaling principle, we vary the event-ready sampling in the same way as in Fig.~\ref{fig:CHSH} and as in Fig.~S3 in \cite{HensenSup}, but we focus on the behavior of the marginal probabilities instead, and on the no-signaling equalities (\ref{NoSignBtoA0}-\ref{NoSignAtoB1})defined above.

\subsubsection{Exhibiting the problem with marginal probabilities}

\begin{figure}[htp]
  \centering
  \includegraphics[width=0.8\linewidth,height=0.8\textheight,keepaspectratio]{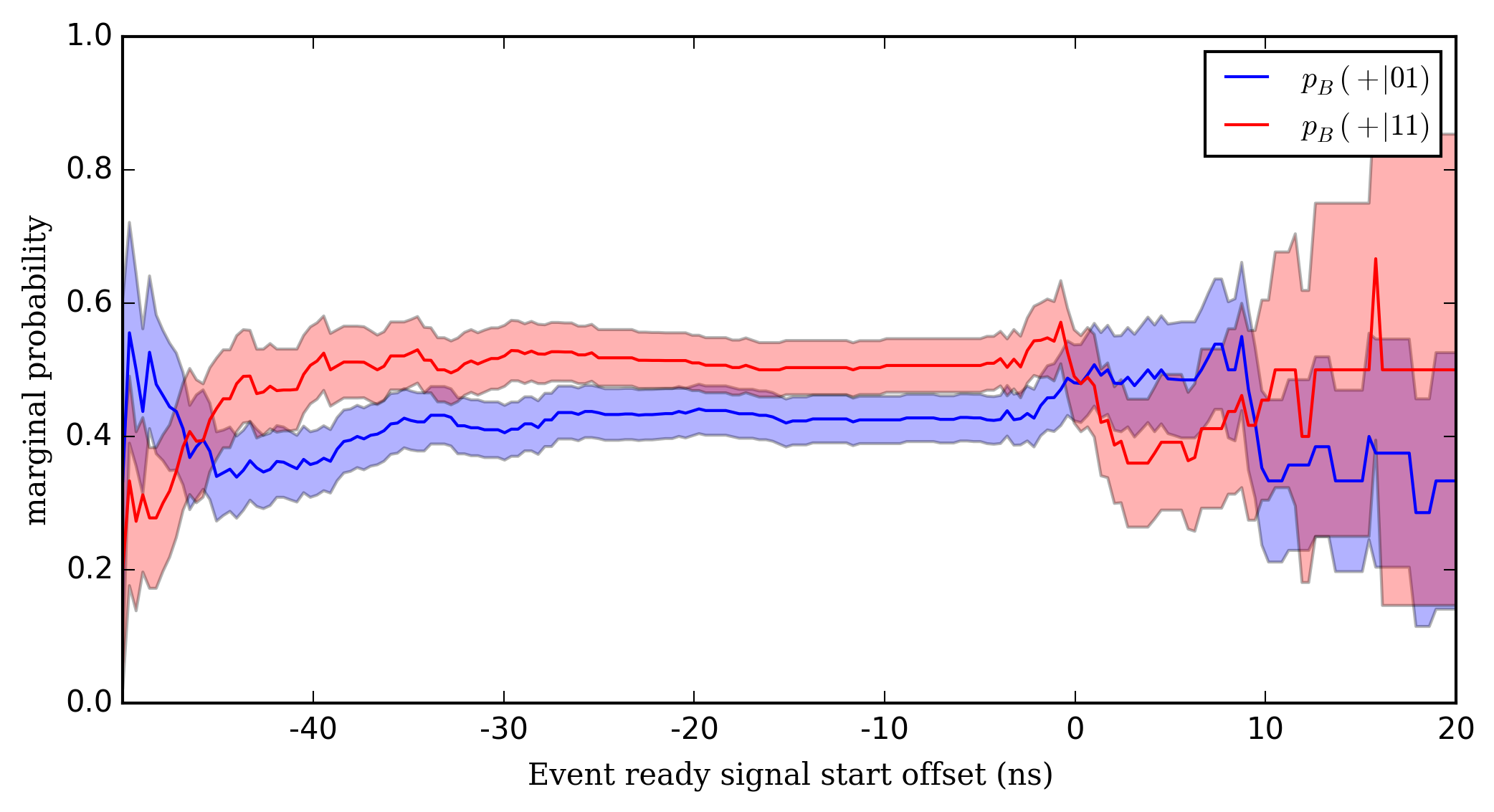}
  \caption{Bob's marginals probabilities for the +1 measurement result, as a function of the window start for the event-ready sampling. The confidence region shown is one standard deviation, calculated according to the conventional analysis.}
  \label{fig:marginals}
\end{figure}

In order to exemplify the nature of the problem, we plot Bob's marginal probabilities for the +1 result when he is measuring along direction 1, as a function of the event-ready window offset (see Fig.~\ref{fig:marginals}). As expressed in Eq.~(\ref{NoSignAtoB1}), the two marginal probabilities associated with different measurement settings for Alice should be equal. They nevertheless differ noticeably when the offset is negative, that is, when the event-ready sample is larger.

Naturally, this effect could simply be due to the finite size of the event-ready sample, in the same way that any finite series of coin tossing will show a deviation from the expected
equal number of heads and tails. We therefore need to estimate the probability that this effect would be due to chance only.

\begin{figure}[htp]
    \centering
  \includegraphics[width=0.8\linewidth,height=0.8\textheight,keepaspectratio]{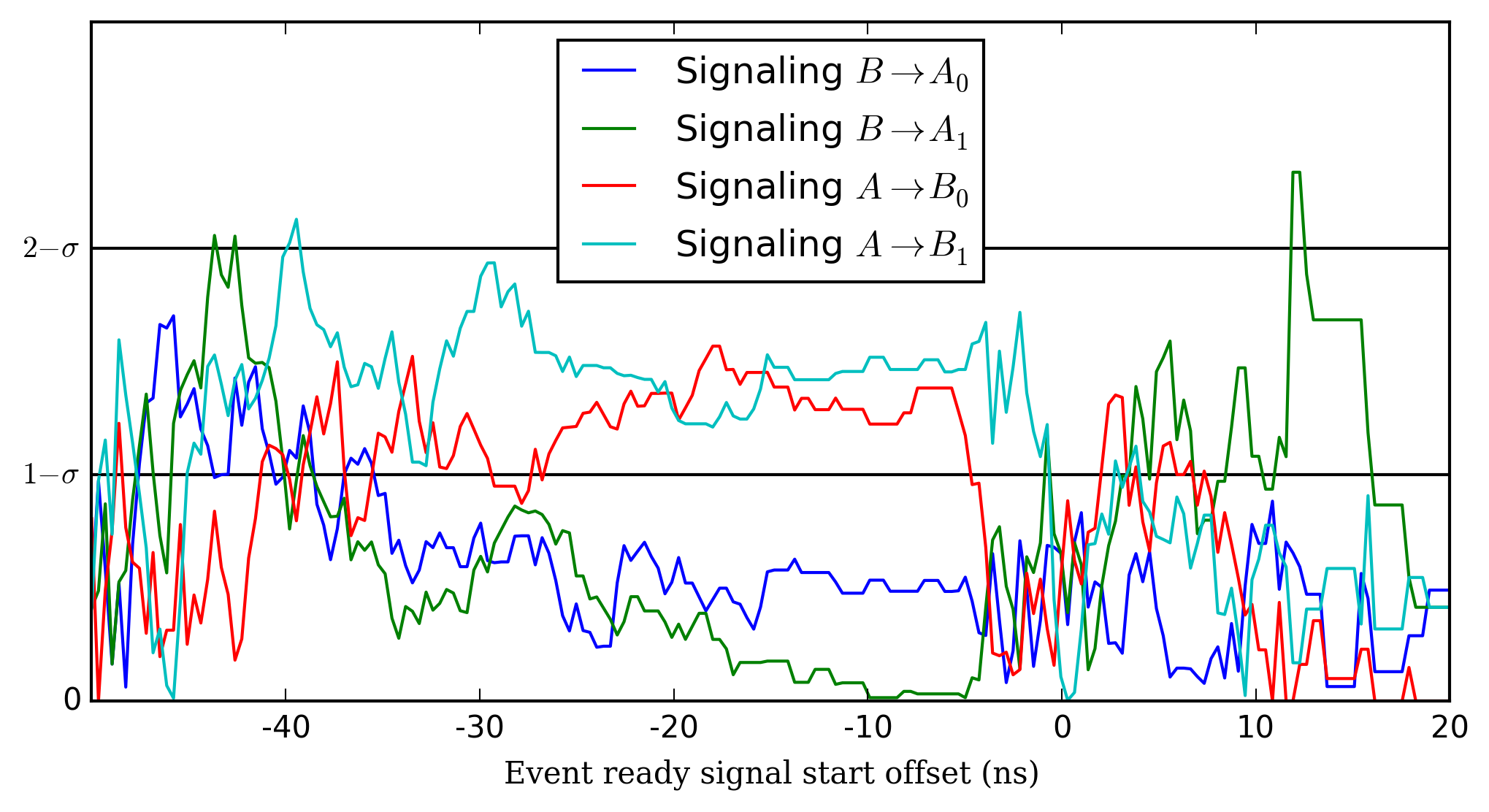}
  \caption{Deviation from the no-signaling equalities for the +1 measurement result as a function of the window start for the event-ready sampling. The deviation remains small from Bob to Alice (except for small sample size), but it becomes larger deviation from Alice to Bob for increased sample size}
    \label{fig:nosign}
\end{figure}

A first step in this direction consists in looking at the behavior of all the marginal probabilities.
Figure~\ref{fig:nosign} shows the magnitude of the violations, in terms of how many standard deviations separates the experiment from each expected no-signaling equality, as a function of the window offset for the event-ready sampling. In the Alice-to-Bob direction, the deviation from no-signaling is consistently larger than one standard deviation for negative offsets (which corresponds to the largest event-ready samples), regardless of which measurement Bob is performing, 0 or 1. For the other no-signaling equalities, in the Bob-to-Alice direction, the experimental observation is less than one standard deviation away from the expectation, except for smaller event-ready samples which are bound to be statistically less significant.

\subsubsection{Increasing further the size of the event-ready sample}

It should be noted that, unlike the violation of local realism, this possible violation of the no-signaling principle is apparently stronger when the sample size is increased, that is, when more coincidences from the unwanted reflections of the laser excitation pulses are included. It would seem to indicate that, if this effect is real, it is not a feature of entanglement between the NV centres, but rather of the excitation pulses.

In order to increase the event-ready sample size further, we propose to modify one additional parameter in the event-ready sampling: the number of previous attempts that are not marked with an invalid marker. This is set to 250 by Hensen \emph{et al} \cite{HensenSup}. We relax this constraints to a value of only 50 instead. It allows to increase the size of the event-ready sampling, at the cost of including in the sample some events that were not recorded with predefined optimal conditions. For instance, it happens if the amount of reflections from the laser excitation pulse becomes too high \cite{HensenSup}.

With this setting, a violation of the CHSH inequality is still observed when all other parameters are kept as set by Hensen \emph{et al}. The size of the event-ready sampling is then $N=335$ (up from $N=245$ in \cite{Hensen}), with the CHSH parameter $S = 2.324$ and a standard deviation $\sigma =0.178$. This is still 1.8 standard deviations away from the local realist bound. It amounts to a $p$-value of 0.035 with the conventional analysis, and 0.055 with the sophisticated analysis of Hensen \emph{et al}.

\begin{figure}[htp]
    \centering
    \includegraphics[width=0.8\linewidth,height=0.8\textheight,keepaspectratio]{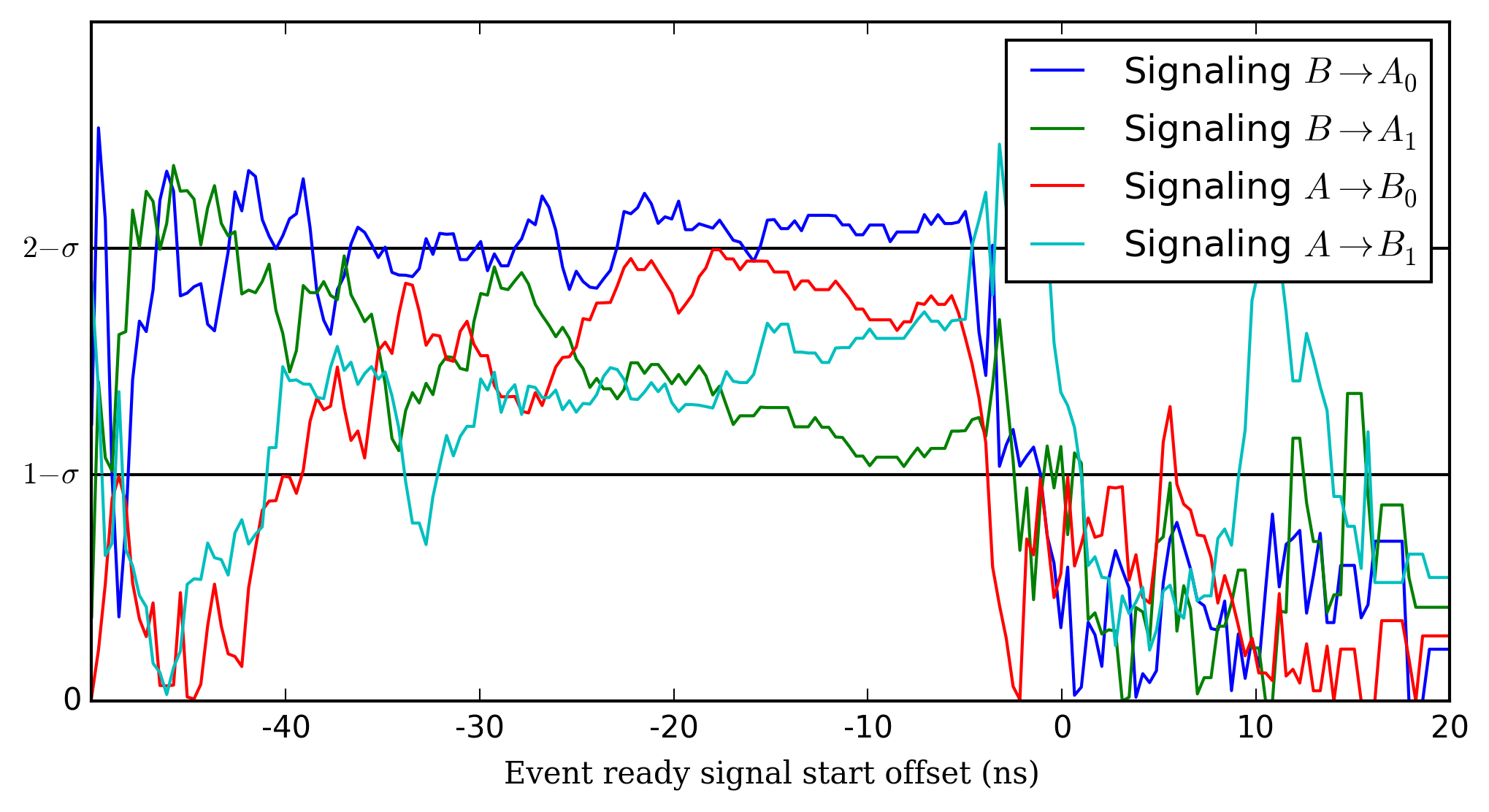}
    \caption{Magnitude of the violation of the no-signaling principle for a larger event-ready sample (with the threshold for invalid markers set at 50), as a function of the window start for the event-ready sampling. All four no-signaling equalities are consistently violated by more than one standard deviation when the sample is the largest.}
    \label{fig:nosignlarger}
\end{figure}

Figure~\ref{fig:nosignlarger} shows that under this condition, all no-signaling equalities of Eqs.~(\ref{NoSignBtoA0}-\ref{NoSignAtoB1}) are consistently violated by one to two standard deviations for negative offsets, that is, when the sample is the largest. Adding the errors in quadrature \cite{Taylor}, this amounts to a combined violation by about 3 standard deviations.

With an event-ready signal start offset set arbitrarily at -20 ns with respect to the value chosen by Hensen \emph{et al}, we reach a sample size of $N=1242$, which is much larger than the 245 events in \cite{Hensen}. The results for the no-signaling equalities and their standard deviations, calculated using the conventional analysis described above, are displayed in Table \ref{Table50}. The resulting four probabilities are quite small, which put the null hypothesis of no-signaling into question. Performing a pooled two-proportion z-test, as in \cite{GiustinaSup}, leads to similar results.

\begin{table}
\begin{center}
\begin{tabular}{lcccc}
\hline
    & $S$ & $\sigma$ & $z$ & $p$\\
\hline
Signaling $\textrm{A}\rightarrow\textrm{B}_0$   & -0.077 & 0.041 & 1.90 & 0.030\\
Signaling $\textrm{A}\rightarrow\textrm{B}_1$   & -0.066 & 0.039 & 1.67 & 0.195\\
Signaling $\textrm{B}\rightarrow\textrm{A}_0$   &  0.086 & 0.040 & 2.17 & 0.058\\
Signaling $\textrm{B}\rightarrow\textrm{A}_1$   &  0.052 & 0.040 & 1.30 & 0.095\\
\hline
\end{tabular}
\caption{Experimental values $S$ of the no-signaling equalities, for a larger event-ready sample ($N=1242$),
with the associated standard deviation $\sigma$, the number $z$ expressing how many standard deviations $\sigma$ separates the experimental observations $S$ from the no-signaling predictions, and the corresponding probabilities $p$ associated with a two-tailed test.}\label{Table50}
\end{center}
\end{table}

\subsection{Is the violation of the no-signaling principle statistically significant?}

In order to understand if this effect is statistically significant, we perform a $\chi^2$ test under the null-hypothesis that the no-signaling principle is independently fulfilled for all four no-signaling equalities of Eqs.~(\ref{NoSignBtoA0}-\ref{NoSignAtoB1}):
\begin{equation}
  \chi^2=\left(\frac{S_{B\rightarrow A_0}}{\sigma_{B\rightarrow A_0}}\right)^2
  +\left(\frac{S_{B\rightarrow A_1}}{\sigma_{B\rightarrow A_1}}\right)^2
  +\left(\frac{S_{A\rightarrow B_0}}{\sigma_{A\rightarrow B_0}}\right)^2
  +\left(\frac{S_{A\rightarrow B_1}}{\sigma_{A\rightarrow B_1}}\right)^2.
\end{equation}
The assumption of independence in natural in the context of a test of the no-signaling principle. First of all, the two signaling directions are in principle independent. A signal going from Alice to Bob's does not necessarily entail that a signal goes from Bob to Alice, and vice versa. Also, the two signaling equalities pointing in the same direction, depend on entirely different joint probabilities. A failure of independence between these joint probabilities would arguably be indicating of a no less serious problem than signaling, which could invalidate the derivation of local realistic bounds \cite{Khrennikov03}.

With the values obtained in Table \ref{Table50}, it amounts to $\chi^2=12.75$. With four degrees of freedom, it means a $p$-value of 0.0125. This low probability raises questions as to the validity of the no-signaling principle in this experiment, but as we will see below, we should nevertheless be careful with the conclusion that we draw. We need to understand if this effect is an artifact specific to a particular choice of event-ready sample, or if it indicates a real issue.

\section{Discussion}

\subsection{Exploring the entire event-ready sample space}\label{entirespace}

Determining if a null hypothesis should be rejected can be a tricky affair, especially when the decision relies on such small sample as with the experimental data at hand. One essential principle in statistics that is particularly relevant in this context is that the rules establishing how the data is collected and how the statistical analysis is performed should be decided independently of the data \cite{Elkouss}. Failure to do so leaves open the possibility to tweak the acquisition or the statistical analysis towards a preferred outcome (the expectation bias \cite{Rosenfeld}). One parameter in particular that should be decided independently of the data is the number of trials to be collected. If the number of trials is not fixed beforehand, then one can be tempted to record trials until the $p$-value reaches desirably low limit: an incorrect practice known as $p$-value fishing \cite{Elkouss,Rosenfeld}.

The best way to enforce this principle is to establish those rules before the data collection starts \cite{Elkouss}, and to strictly adhere to them. Surprisingly, the analysis performed by Hensen et al \cite{Hensen} may have fallen victim of this pitfall. Indeed, the total number of trials that they report is equal to 245. This rather peculiar figure does not suggest that it was decided before the start of the experiment, nor is there any indication in the text that it was. This is quite problematic since the theoretical work that supports the derivation of the expression of the $p$-values used by Hensen \emph{et al} demands that the total number of trials is fixed before the experiment is started \cite{Elkouss,HensenSup}.

\begin{figure}[htp]
    \centering
    \includegraphics[width=0.8\linewidth,height=0.8\textheight,keepaspectratio]{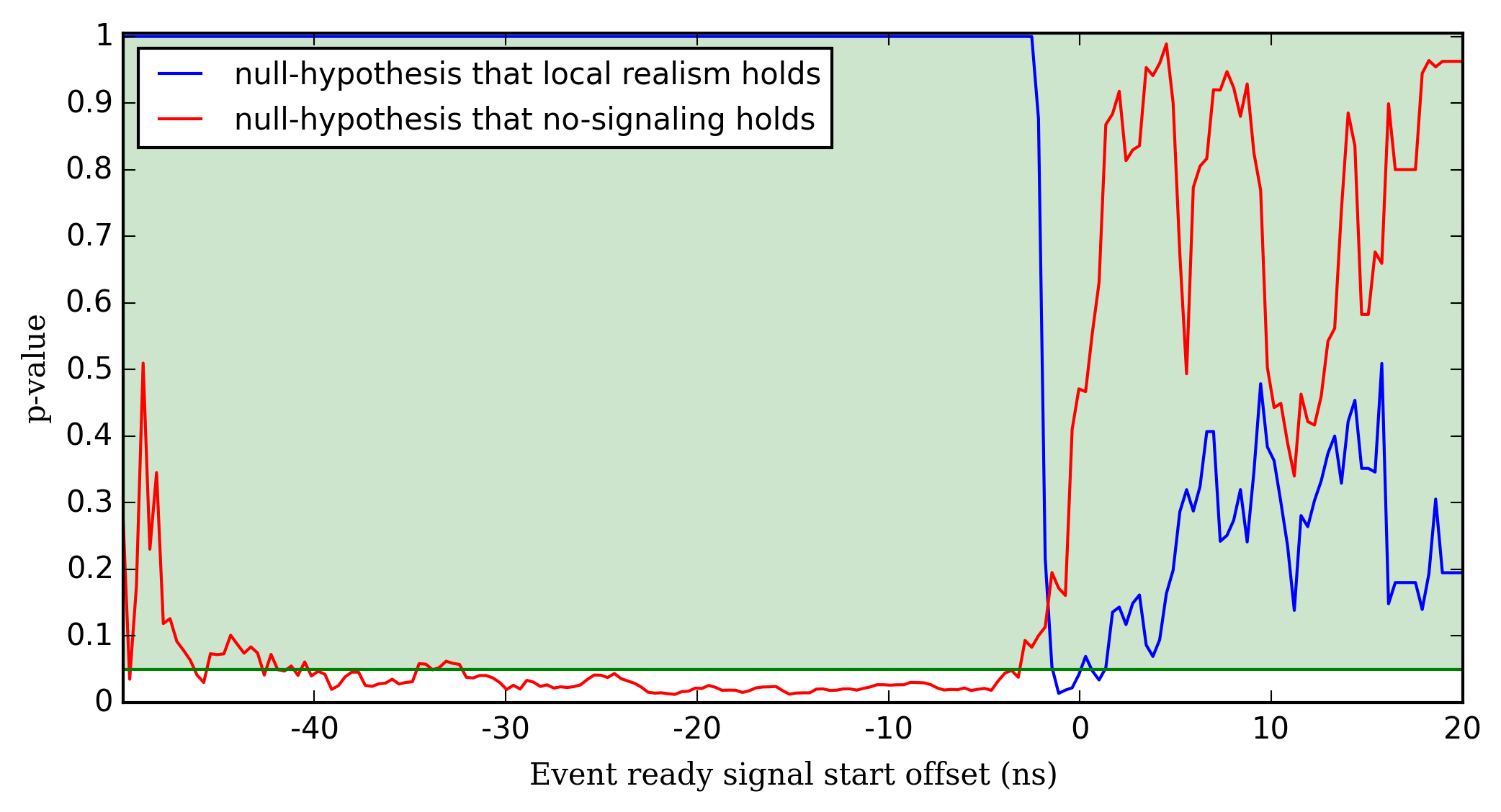}
    \caption{$p$-values for the null-hypotheses of no-signaling and of local realism. It leads to a rejection of local realism (blue line) near the 0 offset, and a rejection of no-signaling (red line) for
a large range of negative offsets.}\label{1Dpvalues}
\end{figure}

Similarly, we need to be careful when we change the event-ready sample selected for analysis in a test of the no-signalling principle. The low $p$-values that we obtained in the previous section could be an artifact specific to the particular event-ready sample that we chose. Since we are not the authors of this experiment, and since we are analyzing the data from an experiment that has long been completed, there obviously is no way that we can decide on the analysis procedure before the experiment is performed. It does not mean that we have to stop right there, otherwise there would be strictly no point for groups to publish experimental data, nor for other groups to analyze these data. However, it certainly means that we have to be careful with the way that we perform the analysis and with our conclusions.

Although we are investigating effects that were not intended to be tested with this data, it should be noted that we do not make any additional hypothesis than those already made while testing local realism. The no-signaling principle is not an hypothesis suggested by the data, but a prediction of both quantum theory and local realist theories. It is therefore perfectly valid to test this hypothesis \emph{a posteriori}, even if such was not the intent of the experimenters.

To avoid the pitfall that we discussed above, while restricting our analysis to a particular event-ready sample, we propose to consider the entire sample space instead, when varying the sampling parameters within limits still producing some data. As we will see, the signaling effect is not specific to a particular sample. It is pervasive, stronger when the sample size is increased significantly, and there is a large range of samples for which it appears statistically significant.

The choice of -20 ns window start offset was admittedly arbitrary, but choosing other negative offsets with a large enough sample gives similar results. There are in fact arguably more event-ready samples showing a significant violation of no-signaling principle than there are showing a significant violation of CHSH, and they are always associated with a larger sample size. This can be illustrated by looking at the $p$-values for both hypotheses when varying the window start offset, as show in Fig.~\ref{1Dpvalues}.

\begin{figure}[htp]
    \centering
    \includegraphics[width=0.8\linewidth,height=0.8\textheight,keepaspectratio]{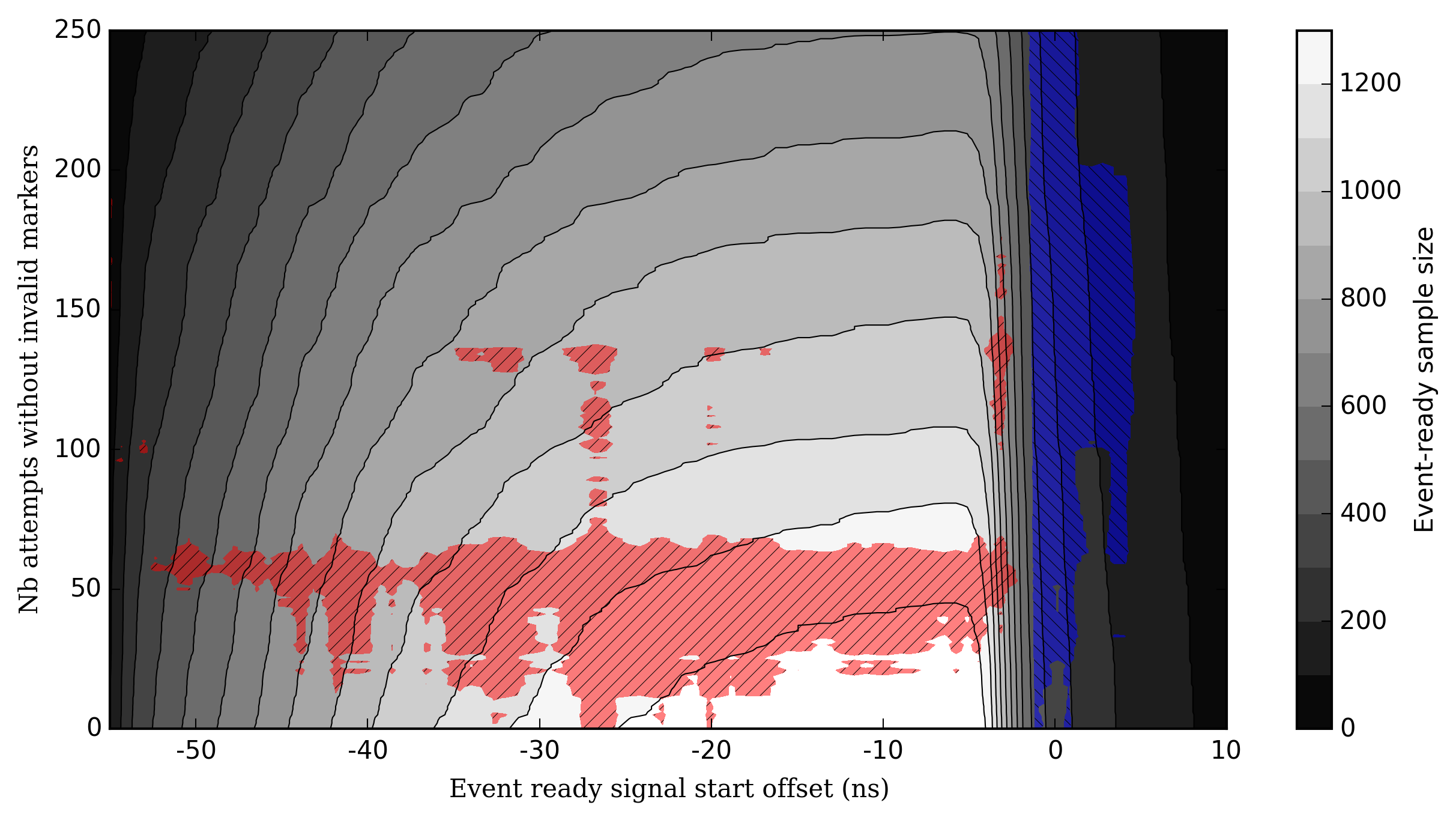}
    \caption{$p$-values for the null-hypotheses of no-signaling and of local realism. The $x$-axis (window offset) and the $y$-axis (invalid markers) scan the event-ready population space. Each point is a possible sample. The blue color contours show the points for which the local realism hypothesis is associated with a $p$-value less than 0.05. The red color contours show the points for which the no-signaling hypothesis is associated with a $p$-value less than 0.05. The black contour (and gray scale) indicate the size of the sample. The point tested in \cite{Hensen} is (0,250).}
    \label{2Dpvalues}
\end{figure}

This can be illustrated further by looking at the $p$-values for both hypotheses when varying the window start offset together with the other sampling parameter that we modified for our analysis, which is the number of previous attempts
that are not marked with an invalid marker (varied from 0 to 250). Note that going further than 250 for this parameter does not change the event-ready samples any more. Figure \ref{2Dpvalues} shows that the violation of the no-signaling principle is statistically significant for a large region of the event-ready population space, and that it happens when the sample size is the largest. By contrast, the violation of local realism (CHSH) is obtained for much smaller sample sizes.

Figure \ref{2Dpvalues} shows the distribution of all $p$-values for the null hypotheses of no-signaling over the entire even-ready sample space, while varying the window offset and the number of invalid markers. The distribution does not appear to be uniformly distributed, as one could expect in the absence of signaling \cite{ShalmSup,Rosenfeld}, but show a sharper tail towards the lowest $p$-values.

\begin{figure}[htp]
    \centering
    \includegraphics[width=0.8\linewidth,height=0.8\textheight,keepaspectratio]{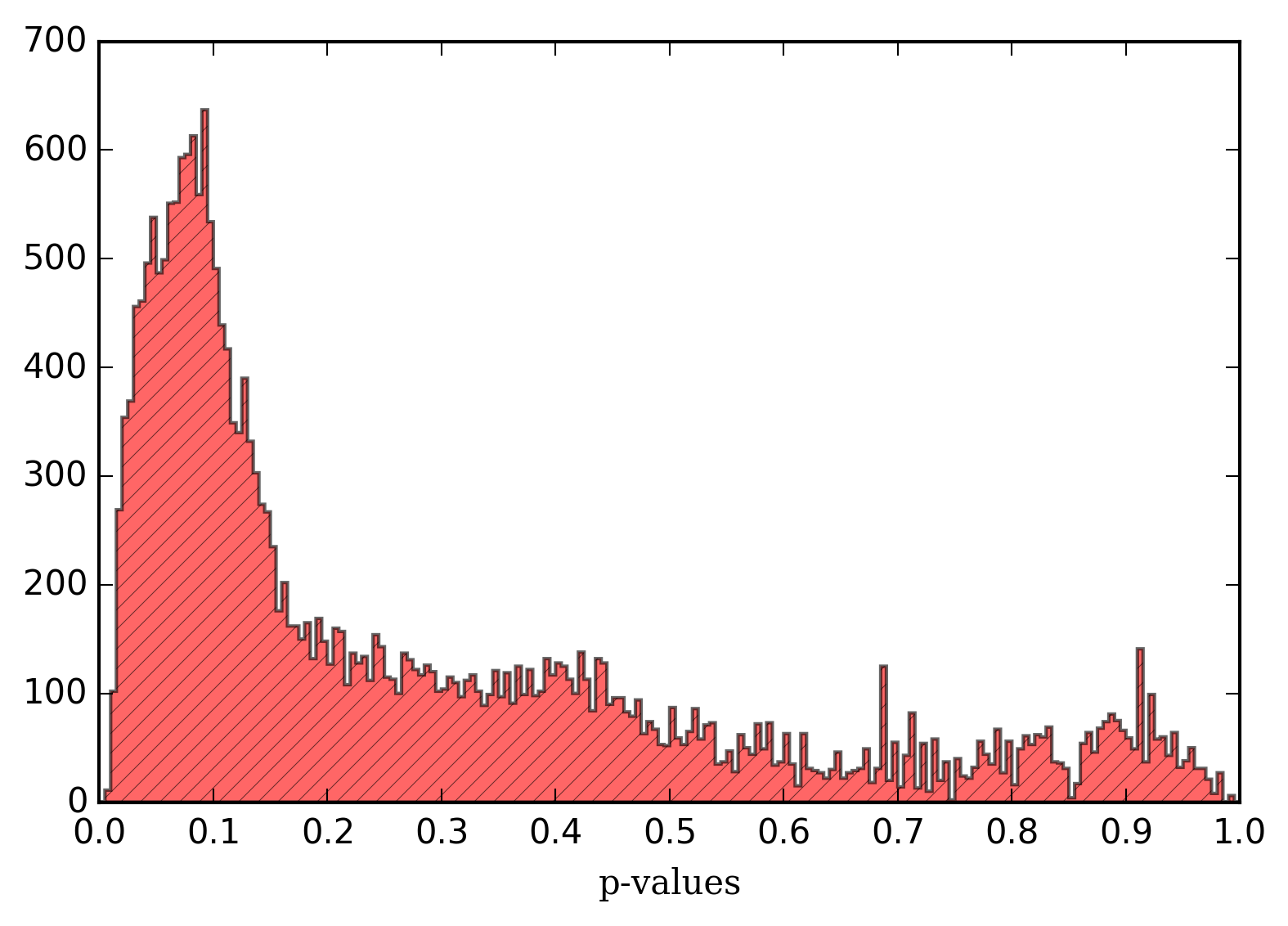}
    \caption{The distribution of $p$-values for the null hypotheses of no-signaling over the entire even-ready sample space. It shows a predominance of low $p$-values that calls for an explanation.}
    \label{Distpvalues}
\end{figure}

\subsection{Relevance of the Gaussian approximation}\label{relevance}

A possible explanation for the signaling effect exhibited in this paper could be that it comes from assuming Gaussian statistics when calculating the $p$-values, and that it would become insignificant with a derivation of $p$-values that does not require this assumption, as in Refs.\cite{Hensen,HensenSup,Elkouss}.

\begin{figure}[htp]
    \centering
    \includegraphics[width=0.8\linewidth,height=0.8\textheight,keepaspectratio]{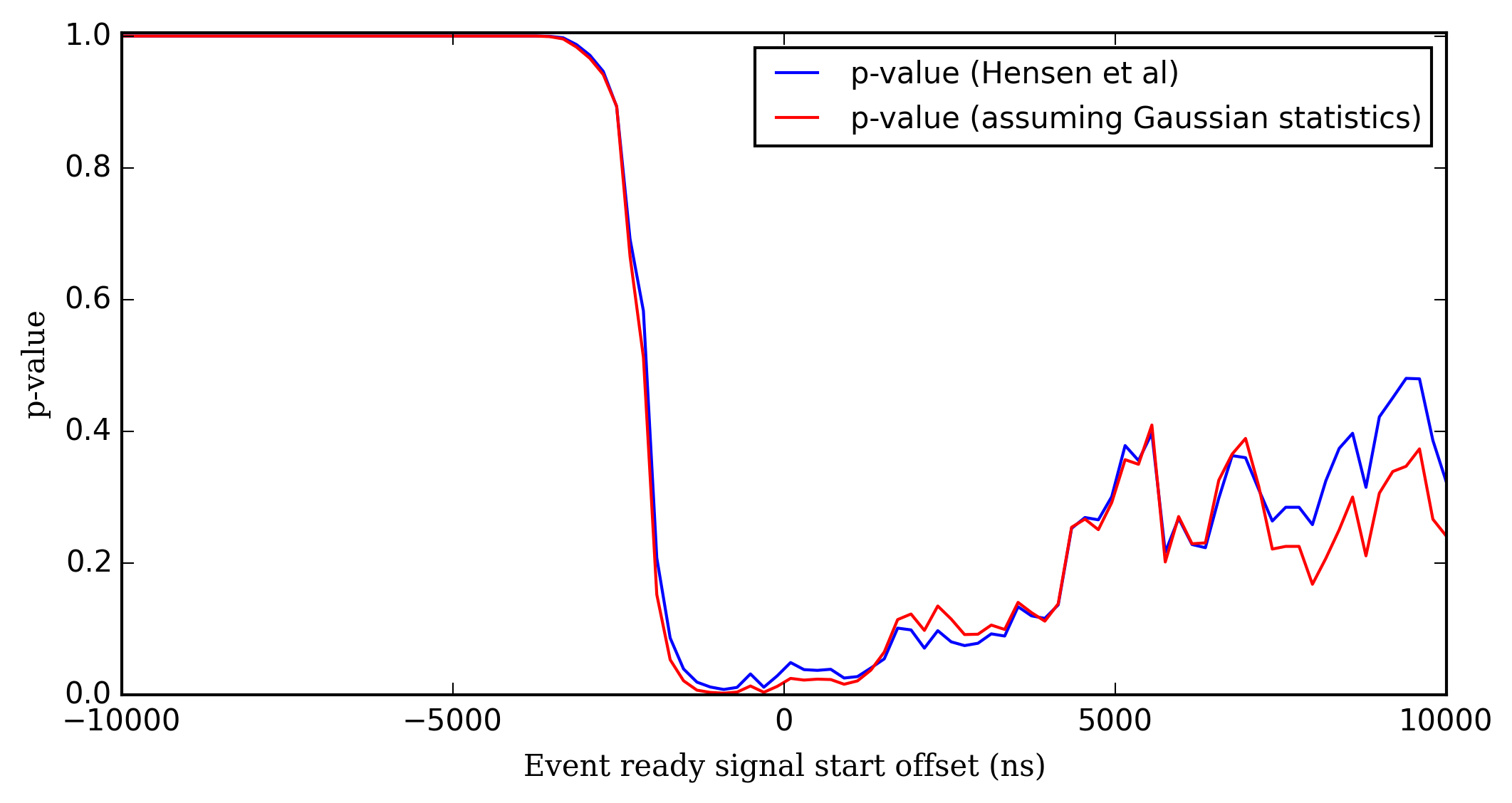}
    \caption{$p$-values comparison for the null hypothesis of local realism, with and without assuming Gaussian statistics. The difference between the two approaches does not appear as particularly meaningful, but rather random, as the event-ready sample is varied.}
    \label{pvaluecompare}
\end{figure}

However, in our view, this would be an unnecessarily complex sophistication, that would only be likely to lead to small differences in the calculated $p$-values \cite{GiustinaSup}. In fact, even in the context of the test of local realism provided by Hensen \emph{et al} \cite{Hensen,HensenSup}, it is unclear that the difference is of any significance.
Comparing the $p$-values using both the conventional approach (assuming Gaussian statistics) and the more sophisticated method not relying on the independent and identically distributed assumption (i.i.d.) \cite{Hensen,HensenSup,Elkouss}, one gets very similar results regardless of which event-ready sampling set is used. This is illustrated in Fig.~\ref{pvaluecompare}, where the $p$-values for the CHSH violation are plotted as the size of the event ready sample is varied, with and without assuming Gaussian statistics. The difference between the two approaches is arguably just noise: sometimes positive, sometimes negative, depending on the sample. It means that having a $p$-value without the memory assumption that about twice as large as with this assumption, for the particular sample chosen by Hensen \emph{et al} \cite{Hensen,HensenSup}, is arguably incidental rather than meaningful.

In this respect, it should be pointed out that with the sophisticated method that makes no assumptions on the memory of the devices \cite{HensenSup,Elkouss}, the best characterisation of the $p$-value is that it is nevertheless bounded from above by the probability that one would obtain assuming i.i.d. directly. The upper bound on the calculated $p$-value is the tail probability of an i.i.d. distribution, even though no i.i.d. assumption was made, a behavior that is not uncommon for a sum of random variables \cite{HensenSup}. Moreover, this bound is sharp \cite{Bierhorst}, which means that the maximal value permitted by the bound can actually be reached by local hidden-variable models that make no assumptions on the memory of the devices. It means that the $p$-value for rejecting local hidden-variable models that permit memory is the same as when restricting to the narrower class of models that rely on the i.i.d. assumption. : allowing for memory does not increase the probability of violating the CHSH inequality under the null hypothesis of local realism \cite{Bierhorst}.

Moreover, even if memory could in principle play a role in an observed violation of Bell inequalities, it arguably bears little relevance to actual physics. So, the added value of using a sophisticated derivation of the $p$-value that is not subjected to this loophole is arguably close to zero.

In any case, since we are not advocating for a ruling out of local realism here, but rather questioning the enforcement of no-signaling, we do not need to rule out the memory loophole, nor any other loophole. If the observed deviations from no-signaling were due to a memory effect, or to any unforseen exotic effect other than signaling, it would be indifferently bad for the experiment. The Gaussian approximation is therefore perfectly sufficient in the case of a test of the no-signaling principle.

\subsection{The Second Experiment}

A second experiment, similar to the first one but with noticeable differences, was performed by the same group five months after the first experiment \cite{HensenSecond}. This second experiment gave a violation of the CHSH inequality by a value of $S=2.35 \pm 0.18$. The results for this second experiment was combined with the first experiment to produce an overall value of $S_\text{combined} = 2.38 \pm 0.136$, yielding a combined $p$-value of $0.0026$ in the conventional approach \cite{HensenSecond}.

In our view, performing a second experiment is perfectly fine if one wants use the results of the first to improve and adjust the parameters, in order to perform a better experiment. However, the idea of combining these experiments into a single one for hypothesis testing in order to obtain a smaller $p$-value is problematic on several account.

Even in the case of a second experiment that would use exactly the same acquisition parameters and analysis procedure as in the first experiment, this is equivalent to having a number of trials that is not fixed before the start of the first experiment. As we have already mentioned in Section \ref{entirespace}, this practice is incorrect and known as $p$-value fishing \cite{Elkouss}. It is also incompatible with the sophisticated method used by Hensen \emph{et al}, since it requires explicitly the number of trials $n$ to be fixed, independently of the observed data \cite{HensenSup,Elkouss}.

As we have already mentioned in Section \ref{entirespace}, this is in fact a fundamental principle of statistical testing: the rules determining  how the statistical analysis is performed should not depend on the data \cite{Elkouss}, and the acquisition procedure, including all acceptance time-windows, should be fixed beforehand \cite{Rosenfeld}.

The trouble is that these rules have been arbitrarily changed between the first and the second experiment. Neither the acquisition parameters nor the procedure for the analysis are the same as in the first experiment: they are different in many ways, and some of these differences are arguably derived from the data (for instance, a larger event-ready window  was used in order to increase the data rate compared to the first experiment, a decision motivated by the data from the first experiment \cite{HensenSecond}).

In the first experiment, the parameters of the event-ready window were determined before the experiment is performed, and then the parameters were kept for the whole duration of the experiment, with a total number of trials equal to 245 for this single event-ready window. By contrast, in the second experiment \cite{HensenSecond}, no less than four different event-ready windows were used, all different from the event-ready window of the first experiment (with number of trials respectively equal to 42, 14, 186 and 58).

The first partition in two separate sets of event-ready windows come from the fact that one detector broke in the course of the second experiment, and had to be replaced. The second partition in two distinct sets comes from the decision to not only measure the $\psi^-$ state as in the first experiment, but also to include as well the measurements of the $\psi^+$ state. The $\psi^+$ state is heralded by two photo-detection events in the same beam-splitter output arm at the event-ready station, instead of in different arms for the $\psi^-$ state. The event-ready windows of the $\psi^+$ state are therefore entirely unlike those of the $\psi^-$ state.

It is quite problematic that these four sets of trials coming from four different event-ready windows are combined \cite{HensenSecond} as if they consisted of a single set with a total number of trials of 300. The decision to include the $\psi^+$ state in the second experiment when it was considered too noisy to be included in the first is particularly striking in this respect. At the very least, each set associated with a distinct event-ready window in the second experiment should have been assigned a distinct $p$-value, and then these $p$-values should have been combined with care (if possible) to produce a global $p$-value for the second experiment. It is also questionable that these four sets are put on some equal footing (except for some weighing) with the first experiment that did consist of a single set of trials with a single event-ready window.

In our view, all these difficulties make it unsuitable to combine these experiments into a single one for hypothesis testing.

\section{Conclusion}

There is certainly something wrong with this observed violation of the no-signaling principle. It just should not happen. Even if this effect could have a perfectly natural cause, for instance due to back reflections of the optical beams, this should have been ruled out by the careful timing of the measurements undertaken by the experimenters \cite{Hensen,HensenSup}.

In all likelihood, the best approach to understand this signaling effect would be to investigate and settle it experimentally. Unfortunately, the possibility of an actual signaling component in the data was not considered by the experimenters, either in the first experiment or in the second, precisely because of the careful timing of all events, as it was considered sufficient to insure independence between the settings and the absence of signaling \cite{HensenSecond}.

Generally speaking, our analysis shows that inferring the fulfillment of the locality condition based on well established theory and careful timing of events is not enough. We would argue that a possible violation of the no-signaling principle should systematically be checked whenever testing for a violation of local realism, as in Refs. \cite{GiustinaSup, ShalmSup, Rosenfeld},
especially for a loophole-free experiment.

\section*{Acknowledgments}

We are grateful to Ivo Pietro Degiovanni for several helpful discussions on the subject, as well as to Richard Gill and Scott Glancy for their insightful comments on a first version of this article. We thank Bas Hensen for providing us the experimental data together with a sample code written in Python.


\begin{thebibliography}{9}
\bibitem{Bell}	J. S. Bell, \textit{Speakable and Unspeakable in Quantum Mechanics}, Cambridge Univ. Press (2004).
\bibitem{CHSH}	J. F. Clauser, M. A. Horne, A. Shimony and R. A. Holt, \emph{Phys. Rev. Lett.} \textbf{23}, 880-884 (1969).
\bibitem{Aspect} A. Aspect, J. Dalibard and G. Roger, {\em Phys. Rev. Lett.} {\bf 49}, 1804-1807 (1982).
\bibitem{Weihs} G. Weihs, T. Jennewein, C. Simon, H. Weinfurther and  A. Zeilinger, {\em Phys. Rev. Lett.} {\bf 81}, 5039 -- 5043 (1998).
\bibitem{Genovese} M. Genovese, {\em Phys. Rep.} {\bf 413}, 319-396 (2005)
\bibitem{GRW} G. C. Ghirardi, A. Rimini, and T. Weber, \textit{Lettere al Nuovo Cimento} {\bf
  27} 10, 293--298 (1980).
\bibitem{Brunner} N. Brunner, D. Cavalcanti, S. Pironio, V. Scarani, S. and Wehner, \textit{Rev. Mod. Phys.} {\bf 86}, 419–478 (2014).
\bibitem{Adenier} G. Adenier and A. Yu. Khrennikov, \textit{J. Phys. B} {\bf 40} 131–141 (2007).
\bibitem{DeRaedt} H. De Raedt, K. Michielsen and F. Jin, \textit{AIP Conf. Proc.} {\bf 1424}, 55 (2012).
\bibitem{GiustinaSup} M. Giustina \textit{et al.}, (Supplementary Information), \textit{Phys. Rev. Lett.}, {\bf 115}, 25 (2015).
\bibitem{ShalmSup}	L. K. Shalm \textit{et al.}, (Supplementary Information), \textit{Phys. Rev. Lett.}, {\bf 115}, 25 (2015).
\bibitem{Rosenfeld} W. Rosenfeld, D. Burchardt, R. Garthoff, K. Redeker, N. Ortegel, M. Rau,and H. Weinfurter, quant-ph arXiv:1611.04604 (2016).
\bibitem{Hensen}  B. Hensen, H. Bernien, A. E. Dreau,	A. Reiserer,	N. Kalb,	M. S. Blok,	J. Ruitenberg,	R. F. L. Vermeulen,	R. N. Schouten,	C. Abellan,	W. Amaya,	V. Pruneri,	M. W. Mitchell,	M. Markham,	D. J. Twitchen,	D. Elkouss,	S. Wehner,	T. H. Taminiau	and R. Hanson,  \textit{Nature}  {\bf 526},   682–686    (2015).
\bibitem{Bednorz} A. Bednorz, quant-ph arXiv:1511.03509 (2015).
\bibitem{Taylor} J. Taylor, \textit{An Introduction to Error Analysis}, University Science Books; 2nd edition (1996).
\bibitem{HensenSup}  B. Hensen, H. Bernien, A.E. Dreau, A. Reiserer, N. Kalb, M.S. Blok, J. Ruitenberg, R.F.L. Vermeulen, R.N. Schouten, C. Abellan, W. Amaya, V. Pruneri, M.W. Mitchell, M. Markham, D.J. Twitchen, D. Elkouss, S. Wehner, T.H. Taminiau, and R. Hanson, (Supplementary Information), \textit{Nature}  {\bf 526},   682–686    (2015).
\bibitem{Khrennikov03} A. Yu. Khrennikov, arXiv:quant-ph/0512178 (2005).
\bibitem{Elkouss} D. Elkouss and S. Wehner, \textit{Npj Quantum Inf.} {\bf 2}, 16026 (2016).
\bibitem{Bierhorst} P. Bierhorst, \textit{Found. Phys.} {\bf 44}, 736–761 (2014).
\bibitem{HensenSecond}	B. Hensen\textit{et al.}, \textit{Sci. Rep.}, {\bf 6} 30289 (2016).
\end{thebibliography}
\end{document}